\documentclass[aps,pra,10pt,twocolumn,showpacs,preprintnumbers,amsmath,amssymb,nofootinbib,superscriptaddress]{revtex4-1}

\usepackage{graphicx}
\usepackage{color}
\usepackage{float}
\usepackage[english]{babel}
\usepackage{blindtext}
\usepackage[T1]{fontenc}
\usepackage{balance}

\newcommand{\MD}[1]{{\color{black} #1}}

\newcommand{\CZG}[1]{{\color{black} #1}}

\newcommand{\csixty}{\text{C}_{60}}
\newcommand{\wcmq}{\text{W}/\text{cm}^2}
\newcommand{\tpulse}{T_\text{pulse}}
\newcommand{\nesc}{N_\text{esc}}
\newcommand{\omlas}{\omega_\text{las}}
\newcommand{\llas}{I_\text{las}}

\newcommand{\bohr}{\mathrm{a}_0}

\newcommand{\grad}{\mathrm{^{\circ}}}

\begin{document}

\title{Strong-field effects in the photo-emission spectrum of the $\csixty$ fullerene}
\author{C.-Z.~Gao}
\affiliation{Universit\'e de Toulouse; UPS; Laboratoire de Physique
             Th\'{e}orique, IRSAMC; F-31062 Toulouse Cedex, France}
\affiliation{CNRS; UMR5152; F-31062 Toulouse Cedex, France}
\author{P.~M. Dinh\footnote{corresponding author :  dinh@irsamc.ups-tlse.fr}}
\affiliation{Universit\'e de Toulouse; UPS; Laboratoire de Physique
             Th\'{e}orique, IRSAMC; F-31062 Toulouse Cedex, France}
\affiliation{CNRS; UMR5152; F-31062 Toulouse Cedex, France}
\author{P.~Kl\"upfel}
\affiliation{Universit\'e de Toulouse; UPS; Laboratoire Collisions-Agr\'egats-R\'eactivit\'e,
IRSAMC; F-31062 Toulouse Cedex, France
}
\author{C.~Meier}
\affiliation{Universit\'e de Toulouse; UPS; Laboratoire Collisions-Agr\'egats-R\'eactivit\'e,
IRSAMC; F-31062 Toulouse Cedex, France
}
\author{P.-G.~Reinhard}
\affiliation{Institut f\"ur Theoretische Physik, Universit\"at Erlangen, Staudtstra\ss e 7, D-91058 Erlangen, Germany
}
\author{E.~Suraud}
\affiliation{Universit\'e de Toulouse; UPS; Laboratoire de Physique
             Th\'{e}orique, IRSAMC; F-31062 Toulouse Cedex, France}
\affiliation{CNRS; UMR5152; F-31062 Toulouse Cedex, France}


\date{draft, \today}

\begin{abstract}
\MD{Considering $\csixty$ as a model system for describing field emission from the extremity of a carbon nanotip,
we explore} electron emission from \MD{this fullerene} excited by an intense, near-infrared, 
few-cycle laser pulse ($10^{13}$-$10^{14}~\wcmq$, 912 nm, 8-cycle).
To this end, we use time-dependent density functional theory augmented by a self-interaction correction. The ionic background of $\csixty$
is described by a soft jellium model. Particular attention is paid to the high energy electrons. Comparing the spectra at different emission angles,  we find that, as a major result of this study, the photoelectrons are strongly peaked along the laser polarization axis forming a highly collimated electron beam in the forward direction, especially for the high energy electrons. \MD{Moreover,} the high-energy plateau cut-off found in the simulations agrees well with estimates from the classical three-step model. \MD{We also investigate the build-up of the high-energy part of a photoelectron spectrum by a time-resolved analysis. In particular,} \CZG{the modulation on the plateau can be interpreted as contributions from intracycle and intercycle interferences.} 
\end{abstract}

\maketitle
\section{Introduction}\label{sec:intro}

Emission from nanometric tips subject to possibly intense laser fields
is of strong current interest as it constitutes a promising source of
a well collimated and coherent beam of ultrafast
electrons~\cite{Hil09}, which can be used in electron diffraction
experiments~\cite{Wil97} as well as electron
microscopy~\cite{Lob05}. Such systems could be especially interesting
in the recollision regime (in which emitted electrons recollide with
the tip in the course of the laser pulse) as this could allow to
generate extremely short electron pulses~\cite{Pau01} and thus pave
the way to sub-femtosecond and sub-nanometric probing of
matter~\cite{Ago04}. Most studies have up to now focused on metallic
nanotips. In particular, laser-induced photo-emission of electrons in
the recollision regime, manifesting itself in the formation of a
plateau at high kinetic energies in the photo-electron spectrum (PES),
has been very recently observed in tungsten tips of na\-no\-me\-tric
size~\cite{Bor10,Sch10,Kru12b}.  In this context, carbon nanotubes
have also been proposed as promising emitting devices, since they
possess an even smaller tip size than the metal tips used so
far~\cite{Car97}. These novel structures have mainly been
characterized by field emission experiments~\cite{Sat97,Nil00}. Due to
this smaller tip size of carbon nanotubes, these systems might thus
lead to even more promising sources for sub-femtosecond /
sub-nanometric probing of dynamic processes in various areas of
physics, chemistry, biochemistry or material science. Only very
recently, laser-assisted photo-electron studies, which have been very
successful in the case of metal nanotips, have been extended to these
novel tips based on carbon nanostructures~\cite{Bio14}.

In relation to this question of field emission from carbon nanotips,
the present work aims at studying to which extent photo-emission by
ultrafast laser pulses can reach the intended recollision regime in
carbon nanostructures. As a starting point, we shall focus our
analysis on the actual extremity of the tip which practically consists
in the cap of a nanotube (the tip is formed by rolling graphene sheets
with various radii and helical structures). The cap
of the carbon nanotubes used in the above mentioned photo-emission
experiments are similar in size and structure to a $\csixty$
cluster~\cite{Sai92}. In this first step we shall thus use $\csixty$
as a test model for the exploratory studies presented here. We
consider the response of $\csixty$ subject to an intense laser
irradiation and characterize this response in terms of photo-electron
spectra (PES).

 The paper is outlined as follows:
Section~\ref{sec:frame} briefly introduces the theoretical framework
and computational details. Section~\ref{sec:results} presents and
discusses the results. Section~\ref{sec:conclusion} completes the
paper with a conclusion.


\section{Formal framework}
\label{sec:frame}

\subsection{Time-dependent local-density approximation}
\label{sec:tdks}

We describe electron dynamics by means of time-dependent density-functional theory at the
level of the time-dependent local-density approximation (TDLDA). 
The LDA is complemented by an average-density self-interaction
correction (SIC), which has been shown to provide a reliable 
theoretical framework of electron dynamics in strong laser fields, 
in particular when the excitation leads to substantial ionization
~\cite{Kaw09,Fen10}. 
The detailed theoretical approach is given elsewhere, and
shall only 
briefly be summarized here. 
The time-dependent single-particle (s.p.)
wave functions $\varphi_{j}(\textbf{r},t)$ are obtained by solving the
time-dependent Kohn-Sham equations~\cite{Run84} (here and in the
  following we use atomic units)~:
\begin{subequations}
\begin{eqnarray}
\label{eq:tdks}
 \mathrm{i}\frac{\partial}{\partial t} \varphi_{j}(\textbf{r},t) 
 &=&
 \{-\frac{1}{2}\nabla^{2}+\upsilon_\mathrm{C}[\rho(\textbf{r},t)]
\cr
 &&  \quad+\ \upsilon_\mathrm{xc}[\rho(\textbf{r},t)]
  +\upsilon_\mathrm{ext}(\textbf{r},t)\} \varphi_{j}(\textbf{r},t),
\\
 \upsilon_\mathrm{ext}
 &=&
 \upsilon_\mathrm{jel}+\upsilon_\mathrm{las},
\end{eqnarray}
\end{subequations}
where $\upsilon_\mathrm{C}$ is the standard Coulomb potential and
$\upsilon_\mathrm{xc}$ the exchange-correlation potential from DFT \cite{Dre90} using the
exchange-functional given in~\cite{Per92}.  
This potential depends on the actual
electron density
\begin{eqnarray}
 \rho(\textbf{r},t)=\sum^{N_{\rm el}}_{j=1}|\varphi_{j}(\textbf{r},t)|^{2}.
\end{eqnarray}
where $N_{\rm el}$ is the number of electrons.  The external one-body
potential $\upsilon_{\rm ext}$ is composed from the potential of the ionic
background potential, here modeled as a jellium (see section
\ref{sec:jellium}), and the potential of the laser field (see section
\ref{sec:las}).

\subsection{Ionic background}
\label{sec:jellium} 
 
The positively charged ionic background is approximated by a jellium
model.  Specifically for $\csixty$ considered here, all carbon ions
are arranged into a shell-like structure~\cite{Kro85}. 
The model for the jellium
potential $\upsilon_\mathrm{jel}$ for the ionic background reads~:
\begin{subequations}
\label{eq:psrho}
\begin{eqnarray}
  \upsilon_\mathrm{jel} (\mathbf r)
  &=&
  -\int \textrm d^{3} \mathbf r'
  \frac{ \rho_\mathrm{jel}(|\textbf{r}'|) } {|\textbf{r}-\textbf{r}'|}
  +
  \upsilon_\mathrm{ps}(|\mathbf r|)
  \;,
\label{eq:vjel}
\\
  \rho_\mathrm{jel} (r)
  &=&
  \rho_0 \,g(r)
  \;,
\label{eq:pspot}
\\
 \upsilon_\mathrm{ps} (r)
  &=&
  \upsilon_0 \, g(r)
  \;,
\label{eq:WS}
\\
   g(r)
   &=&
   \frac{1}{ 1+\exp\left[(r\!-\! R_-)/\sigma\right] }\,
   \frac{1}{ 1+\exp\left[(R_+\!-\!r)/\sigma \right] }
   \;,
\label{eq:gfunc}
\\
   R_\pm
    &=&
    R\pm\frac{\Delta R}{2}
    \;.
\label{eq:Rpm}
\end{eqnarray}
\end{subequations}     
The jellium density $\rho_\mathrm{jel}$ is modeled by a sphere of
positive charge with a void at the center \cite{Pus93,Bau01,Cor03}.
The Woods-Saxon profile $g$ generates a transition from bulk shell to
the vacuum (inside and outside), providing soft surfaces.  
Furthermore, we employ a pseudo-potential
$\upsilon_\mathrm{ps}$
 in addition to the potential created by the
jellium density, as proposed in Ref.~\cite{Rei13}, which is tuned to
ensure reasonable values of the single-particle energies.  The average
radius $R$ of the jellium cage is taken from experimental data as
$R=6.7~\bohr$~\cite{Hed91}.  The thickness of the jellium shell
$\Delta R$, the surface softness $\sigma$, and the depth of the
potential well $\upsilon_0$ are adjustable parameters for which we use
here $\sigma=0.6~\bohr$, $\upsilon_0=1.9~\mathrm{Ry}$, and $\Delta R=
2.57~\bohr$. The bulk density $\rho_0$ is determined such that
$\int\textrm{d}^3\mathbf{r}\,\rho_\mathrm{jel}
(\mathbf{r})~=N_\mathrm{el}=238$.
Note that this number of electrons is different from 240 for a real
$\csixty$, but no jellium model is capable to 
place the electronic shell closure at this value  so far
(unless one uses a deliberate modification of the occupation numbers
\cite{Mad08}). 
Most jellium models for $\csixty$ have the shell
closure at $N_\mathrm{el}=250$ \cite{Pus93,Bau01,Cor03}. The present
model with soft surfaces comes to $N_\mathrm{el}=238$ which is much
closer to reality. By virtue of the choice of model parameters, 
the electronic properties of $\csixty$ are well reproduced, 
in particular the 
ionization potential (IP) at $E_\mathrm{IP}=0.56$ Ry, that is identical to the
experimental value~\cite{Lic91b}, a HOMO-LUMO gap of 0.14 Ry, which 
is well within
the range of the experimental values 
(0.12-0.15~Ry~\cite{Sat10}), as well as a good 
description of the photo-absorption
spectrum (for details, see~\cite{Rei13}).

The use of the jellium approximation, nevertheless, 
requires some words of
caution. The standard procedure is to use a detailed ionic
background coupled to the electrons through pseudo-potentials.
There exist numerous investigations for
$\csixty$~\cite{Kor10,Tof10} and other carbon nanostructures
such as graphene~\cite{Ara04} or nanotubes~\cite{Dri11}.  An
elaborate description of $\csixty$ with an involved orientation
averaging is needed for
detailed observables such as photo-electron spectra (PES) 
and photo-angular distributions (PAD) \cite{Wop15,Gao15,Wop15b}. 
A key issue in the
present investigations is the ponderomotive motion of the
electron in the laser field associated with huge excursions of the
electron~\cite{Her12}.  This requires extremely large simulation
boxes which become unaffordable for a grid representation in full
3D. The jellium model, together with the linearly polarized laser
field, has cylindrical symmetry. This allows us to use a
cylindrical (2 dimensional) box which renders the necessary
huge boxes
feasible. As an additional benefit, we can also compute 
angular distributions
without the extra expense of orientation
averaging \cite{Wop10a,Wop10b}. As far as the jellium model 
is concerned, 
it is a powerful
approximation as it provides an appropriate description of many
features of the electronic structure and dynamics in solids
\cite{Ash76}, cluster physics \cite{Eka84,Bra93}, and $\csixty$
\cite{Bau01}.  However, two aspects have been
sacrificed.  The first one is that the returning electron
collides with the jellium well instead of with a carbon
ion. Although the potential of the ionic background is very 
steep, we are probably underestimating the actual yield of
high-energy electrons. The second
aspect is that we ignore ionic motion. 
As a consequence, we miss effects from phonon 
coupling~\cite{Gun95a} as well
as from electronic dissipation \cite{Rei15a}.
However, since we are using femtosecond laser pulses, 
their effect should not drastically affect the main findings 
presented here.

\subsection{Laser field}
\label{sec:las}

The laser field is taken to be linearly polarized 
along the $z$ direction, with a sin$^2$-shaped envelope,
\begin{equation}
  \mathbf E_\mathrm{las} (t) 
  =
  E_0\sin^2 \left(\pi t/\tpulse\right)\sin(\omlas t+\phi_\mathrm{CEP})\,\mathbf{e}_z
  \;.
\label{eq:laser}
\end{equation}
Within the dipole approximation in length gauge, the laser-electron 
interaction is given by 
$V_\mathrm{las}(\mathbf{r},t)=-\mathbf{E}_\mathrm{las}(t)
\cdot\mathbf{r}$. 
We use a laser frequency $\omlas$=1.36 eV (=912 nm) and a total
 duration $\tpulse=24$~fs.  The laser strength $E_0$ 
is varied from 0.0113~V/$\bohr$ to
0.0453~V/$\bohr$, corresponding to laser intensities 
$I_\mathrm{las}$ from $10^{13}~\wcmq$ to $1.6\times 10^{14}~\wcmq$.  It is
instructive to characterize these laser conditions in terms of the
Keldysh parameter
$  \gamma \displaystyle
  =
  \sqrt{2E_\mathrm{IP}}\, \omlas/E_0\,
$ where $E_{\rm IP}$ is the ionization potential~\cite{Kel65}.
The current combination of $\omlas$ and $E_0$ spans the interval
$0.55\leq\gamma\leq$ 2.2, i.e., from multi-photon ionization for
$\gamma>1$ to tunneling ionization for $\gamma<1$. This transition has
been experimentally studied in PES of rare gases~\cite{Mev93}.
In the above expression, $\phi_\mathrm{CEP}$ is the carrier-envelope phase (CEP).
A recent combined experimental/theoretical study on strong-field ionization in $\csixty$~\cite{Li15} reported a remarkable CEP effect. 
In these studies, a pulse duration of 4 fs and a central frequency of $\omega_\mathrm{las}=1.72$ eV have been used. However, in the present work, 
much longer pulses (of about 8 optical cycles) are used. \MD{We have performed a} systematic analysis by varying the CEP \MD{and found} no
significant influence on the PES. As a consequence, only results for $\phi_\mathrm{CEP}=0$ are shown below. 


\subsection{Numerical representation}
\label{sec:grid}

A detailed description of the numerical treatment can be found in
\cite{Cal00,Rei03a}. Here, we give a brief account and 
specify the actual
numerical parameters used.  Wave functions, densities, and
fields are represented on a cylindrical grid in coordinate space.
As already mentioned, a major issue of the present investigation
is the rescattering of electrons which requires very large
  computational boxes for a complete description of the huge electron
  excursions.  We have thus made systematic investigations on the
  impact of box parameters.
The final choice is a compromise between acceptable numerical
  cost, accuracy and robustness. The chosen dimensions of the
numerical box are 500~$\bohr$ along the laser polarization direction
($z$ axis) and 250~$\bohr$ in radial direction ($r$ coordinate). The
grid spacing is taken to be $0.5~\bohr$, which allows us to represent
kinetic energies up $\sim$140 eV. 
The electronic ground state
is determined by the damped gradient method~\cite{Rei82}. 
The Kohn-Sham wave
functions are propagated in time using the time-splitting
technique~\cite{Fei82}, and absorbing boundary conditions are
used to remove all (emitted) electrons which have reached the
boundaries of the box.  They consist of 70 grid points (=35
a$_0$) at each of the margins.

\subsection{Observables}

We have studied various observables related to the response of the system, in particular to ionization.  
The total ionization is
calculated as the difference between the initial number of
electrons $N_{\rm el}$ and those left in the box at a given time $t$~:
\begin{equation}
  N_\mathrm{esc}(t)
  =
  N_\mathrm{el}-\displaystyle\int\textrm{d}^3\mathbf{r}\,\rho(\mathbf{r},t). 
\label{eq:nesc}
\end{equation}
This quantity gives an indication on the charge state of
$\csixty$ after irradiation~\cite{Fen10}.  Another observable is the
electronic dipole moment
\begin{equation}
  \mathbf{D}(t)
  =
  \displaystyle\int\mathrm{d}^3\mathbf{r}\,\rho(\mathbf{r},t)\,\mathbf{r},
\label{eq:dip}
\end{equation}
which characterizes the electronic response in time.  It is
mostly used to compute photo-absorption spectra using
spectral analysis \cite{Cal97b}. Here we use it in the time domain
to visualize the electron dynamics of the system.  In the following,
we will consider particularly the dipole moment parallel to the
laser polarization, that is $D_z$.

Most importantly, we will concentrate our analysis of electron
emission on the angular-resolved photo-electron spectra (ARPES)
  yield $\mathcal{Y}(E_\mathrm{kin},\theta)$, that is the yield of
asymptotic kinetic energies $E_\mathrm{kin}$ of electrons emitted in
direction of angle $\theta$. 
To evaluate $\mathcal{Y}$ from our TDDFT simulations, we employ
the method initiated in \cite{Poh00a,Poh01} and extended in \cite{Din13} to the case of
  strong fields as used here.
In brief, the PES is
computed by recording the single-particle wave functions
$\psi_j(t,\mathbf{r}_{\mathcal{M}})$, $j=1, \ldots, N_\mathrm{el}$, at
selected sampling points $\mathbf{r}_{\mathcal{M}}$ near the absorbing
boundary. Once the simulation is completed,
one computes the Fourier transform from time to frequency domain
$\widetilde{\psi_j} (E_\mathrm{kin},\mathbf r_{\mathcal M})$
augmented by a phase factor accounting for the external
field ~\cite{Din13}. The PES then reads
\begin{equation}
  \mathcal{Y}(E_\mathrm{kin},\theta) 
  \propto 
  \sum_{j=1}^{N_{\mathrm{el}}}
 \lvert \widetilde{\psi_{j}}(E_\mathrm{kin},\mathbf r_{\mathcal M})
 \rvert^2 \quad.
\label{eq:pes}
\end{equation}
The sampling points are chosen to cover a mesh of emission angles
$\theta$.  This angle $\theta$ at detection point
$\mathbf{r}_{\mathcal{M}}$ is defined with respect to the laser
polarization $\mathbf e_z$~: forward and backward emissions
correspond to $\theta=0^{\circ}$ and $\theta=180^\circ$
respectively. 
The energy and angle resolution for the PES is 0.04 eV and
$1^{\circ}$ respectively.

\section{Results and Discussions}
\label{sec:results}

\subsection{Angular dependence of the photoelectron spectra } 
\label{sec:angle}

\begin{figure}[htbp]
 \centering
 \includegraphics[width=\linewidth]{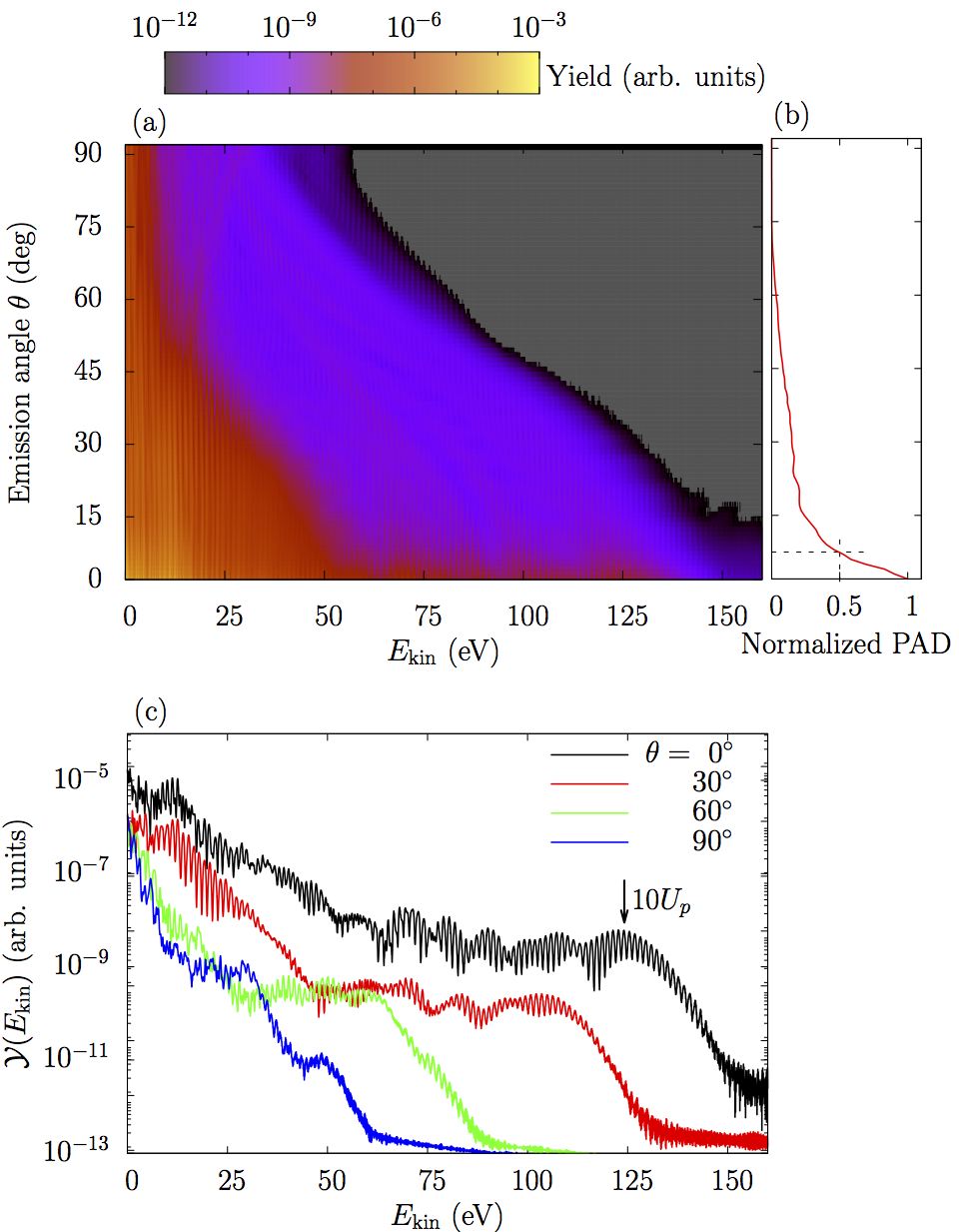}
 \caption{\label{fig:pespad} (a): Color map of angle-resolved photo-electron spectra (ARPES)
as a function of kinetic energy and emission angle, in logarithmic scale for
   the yield, from
   $\csixty$ excited by a laser pulse with following parameters~: $\omlas=1.36$~eV,
   $T_{\rm pulse}=24$~fs, $I_\mathrm{las}=1.6\times{10}^{14}~\wcmq$.  (b): PAD obtained from integration of the ARPES
over the high-energy range of [50--160 eV], normalized to 1 at $0\grad$. The
   dashed line marks the cone angle at full width at half maximum.  (c): PES for emission angles at $\theta =
   0\grad, 30\grad, 60\grad, \rm{and}~90 \grad$ (in logarithmic
   scale), with an arrow indicating the $10U_p$ cutoff energy.}
\end{figure}

The upper panel of Fig.~\ref{fig:pespad} shows the full ARPES at intensity
$I_\mathrm{las}=1.6\times{10}^{14}~\wcmq$ complemented by 
the angular distribution (PAD) on the right. 
This PAD is obtained from integration 
of the ARPES over the kinetic energy interval
50--160~eV, while the lower panel shows the PES for selected
angles, as indicated. 
The ARPES clearly indicates that electron emission is more pronounced for higher energies and
is strongly focused in forward direction ($\theta=0\grad$)~:
The ratio of the yield at $0\grad$ and at $90\grad$ increases from $\sim100$
at low energies to $\sim10^5$ in the high energy range (100--150 eV).
Analyzing the angular distributions in Fig.~\ref{fig:pespad} (b) yields a focusing of $\sim{7}\grad$ (full width at half maximum), similar to values observed in gold nanotips ~\cite{Par12}.

In Fig.~\ref{fig:pespad}(c), we present the photoelectron 
spectra for different emission angles, as indicated.  
All four PES start with a nearly exponential decrease and then,
for higher energies, develop into a broad plateau which extends
up to a distinct cut-off. \MD{We will discuss this structure in more details in the next section. Note that} 
this \MD{typical pattern} is well known 
and can be qualitatively understood by the 
so called ``three-step model\rq{}\rq{}~\cite{Cor93}: 
the electron is ionized during the laser field through 
tunnel ionization, then accelerated in the electric 
field, driven back to the ion core and gains a large
amount of energy through recollision with the latter. 
Within this simple picture, the cut-off appears at
$\sim10 U_p$ with $U_p=\frac{I_\mathrm{las}}{4\, \omlas^{2}}$ being the ponderomotive energy.
\MD{Here,} \CZG{the angular dependence of the cut-off roughly 
follows a $\cos \theta$ dependence, similar to the angular tendency of cut-off shown 
in~\cite{Pau94,Cor08}}. A further interesting finding is the 
oscillatory structure of the PES within the high energy plateau. 
Similar structures have been observed in 
other systems~\cite{Kop99,Fro09}, where they were interpreted as interference 
effects of different electron trajectories leading to the same final kinetic energy. 
In Sec. C, we will analyze the time evolution of the spectra, and we will show that this interpretation is consistent with our results. In the following section, we will first address the intensity dependence of the photoelectron spectra.  

\subsection{Impact of laser intensity}
\label{sec:impactlas}

\subsubsection{PES as a function of laser intensity}

\begin{figure}[htbp]
 \centering
 \includegraphics[width=0.92\linewidth]{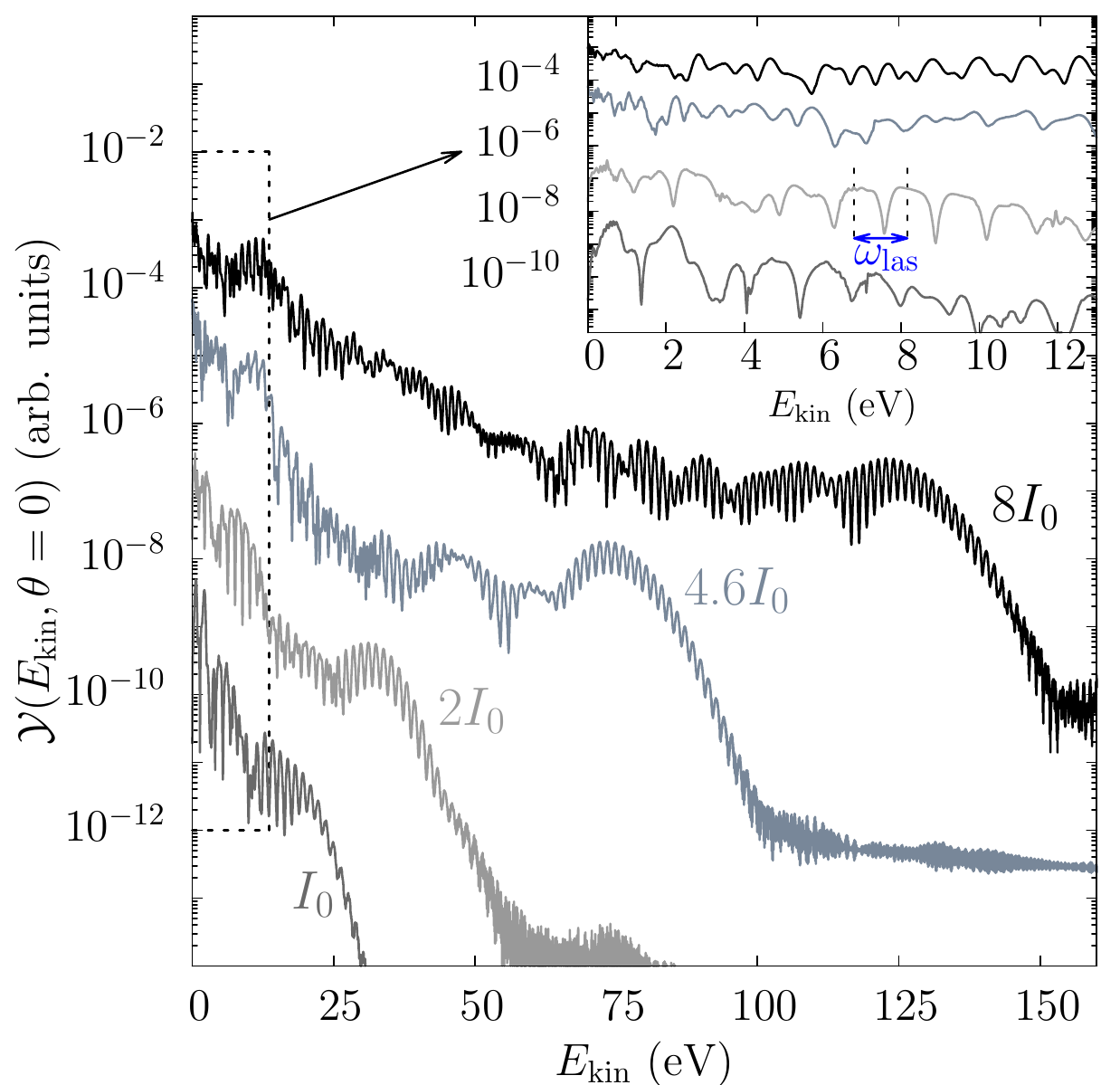}
 \caption{\label{fig:PES_Intens} Samples of
   photo-electron spectra in forward direction 
   in $\csixty$ irradiated by laser pulses with 
   $\omlas=1.36$~eV (912~nm), $T_{\rm pulse}=24$~fs, 
and four intensities as indicated ($I_0=2\times 10^{13}~\wcmq$). The curves are vertically shifted
   for a better graphical discrimination.
Inset~: zoom 
   in the region $0<E_\mathrm{kin}<13$ eV contributed by 
   direct emissions. }
\end{figure}

To study the influence of the laser intensity on PES of
  irradiated $\csixty$, four intensities have been explored, namely
$I_0$, 2$I_0$, 4.6$I_0$, and $8I_0$ where $I_0=2\times10^{13}~\wcmq$,
corresponding to values of the Keldysh parameter $\gamma=1.5$, 1.1,
0.7, and 0.5 respectively.  Figure~\ref{fig:PES_Intens} collects the
results for the PES in forward direction ($\theta=0 \grad$).  
They
exhibit similar pattern at all four considered intensities~: on the
low-energy side, they show a more or less extended plateau (see 
inset), then turn to a rapid exponential decrease, and finally
develop a broad second plateau which extends to high energies.  Both
plateaus are shifted towards higher kinetic energy with increasing
laser intensity, whereby the width of the plateau increases as well
and thus the upper energy cut-off of the second plateau increases accordingly.

We first concentrate on the PES at low energies ($<13$~eV)
  magnified in the inset of Fig.~\ref{fig:PES_Intens}.  For the two
lowest intensities ($I_0$ and $2I_0$), they
show a dense sequence of peaks which are separated by
the photon energy $\omlas=1.36~\rm{eV}$.  This is the typical
pattern for Above-Threshold Ionization (ATI), which has been well
studied for $\csixty$ experimentally~\cite{Cam00,Kje10} and
theoretically~\cite{Poh04a,Wop15,Gao15}.  These equi-spaced
patterns are washed out when the laser intensity increases to
the two highest values ($4.6\,I_0$, $8\,I_0$).  
\CZG{This 
can have several reasons: first, due to the significant ionization 
at these intensities, the cluster progressively becomes charged during the laser 
pulse, leading to changes in the electronic structure. This 
effect has been observed in sodium clusters, see e.g.~\cite{Poh00a,Poh04a}, 
and preliminary investigations have clearly shown similar effects  
in the present case too. However, 
additional blurring due to other effects, 
like the space charge, may also contribute, and will be
analyzed in future studies}.

\subsubsection{Analysis of  the high energy plateau}

As mentioned above, the high-energy plateau is generated by strong-field ionization (SFI)
mechanism, and the characteristic cut-offs can be found 
approximately at
$10.007~U_p$~\cite{Pau94}.  Moreover, the derivation of a
semiclassical cut-off law~\cite{Bus06}, based on the Strong Field
Approximation (SFA), reveals that the IP also plays a role in the SFI
regime, and can be estimated by~\cite{Bus06}:
\begin{equation}
  E_\mathrm{cut}^\mathrm{(SFA)}
  =
  10.007 \, U_p+0.538 \, E_\mathrm{IP}
  \;,\
  U_p
  =
  \frac{I_\mathrm{las}}{4\, \omlas^{2}}
  \;,
\label{eq:ecut}
\end{equation}  
Because of the inverse quadratic dependence of $U_p$ on the 
laser frequency $\omlas$, the low value of 1.36~eV used here delivers a large
ponderomotive energy and large cut-offs in the PES. For instance, 
at 4.6$I_0$, we have $U_p$=7.1 eV, largely exceeding the photon 
energy and  being even comparable to $E_\mathrm{IP}$, 
and $E_\mathrm{cut}^\mathrm{(SFA)}=75$ eV in Fig.~\ref{fig:PES_Intens}.

\begin{figure}[hbtp]
 \centering
 \includegraphics[width=\linewidth]{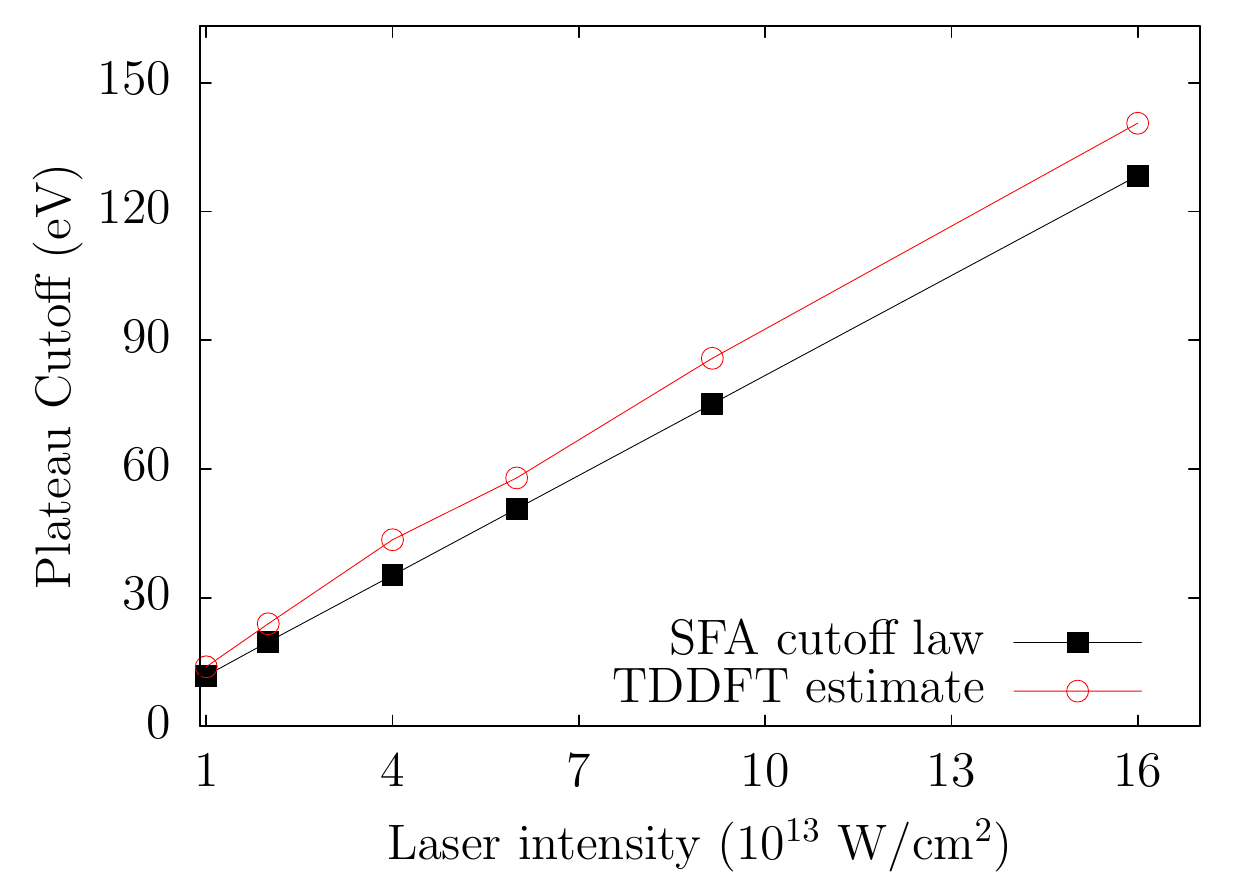}
 \caption{\label{fig:cutpos} Cutoff energies
   of the high-energy plateau in the PES as functions of
   the laser intensity, from TDLDA calculations 
   compared with the estimate (\ref{eq:ecut}) deduced from the strong field
   approximation (SFA).}
\end{figure}

We have extracted the energy cut-off from our calculated PES, 
denoted by
$E_{\rm cut}^{({\rm TDLDA})}$ to compare it to 
the simple estimate $E_{\rm cut}^{({\rm SFA})}$ given by Eq.~(\ref{eq:ecut}).
More precisely, for a given PES, $E_{\rm cut}^{({\rm TDLDA})}$ is obtained as the intersect
of two
fitting exponential curves at each side of the plateau, similarly to
the procedure applied on experimental data~\cite{Kru11}.  
The comparison of $E_{\rm cut}^{({\rm TDLDA})}$ and $E_{\rm cut}^{({\rm SFA})}$ is presented
in Fig.~\ref{fig:cutpos}.
We first note that both energy cut-offs grow linearly with laser intensity, pointing towards the validity 
of the simple classical scaling law.
However, our numerical simulations show a slightly steeper slope.
If one interprets the numerical results obtained by the 
TDLDA approach in the light of the simple classical scaling 
law (\ref{eq:ecut}), the difference may be explained by a field 
enhancement of about 5 to $10\%$. Further investigations are
needed to confirm this interpretation. Note however that
a similar effect has been observed
in metallic nanotips~\cite{Sch10,Par12,Kru12a}. 

\subsubsection{Ponderomotive oscillations}

\begin{figure*}[htbp]
 \centering
 \includegraphics[width=\linewidth]{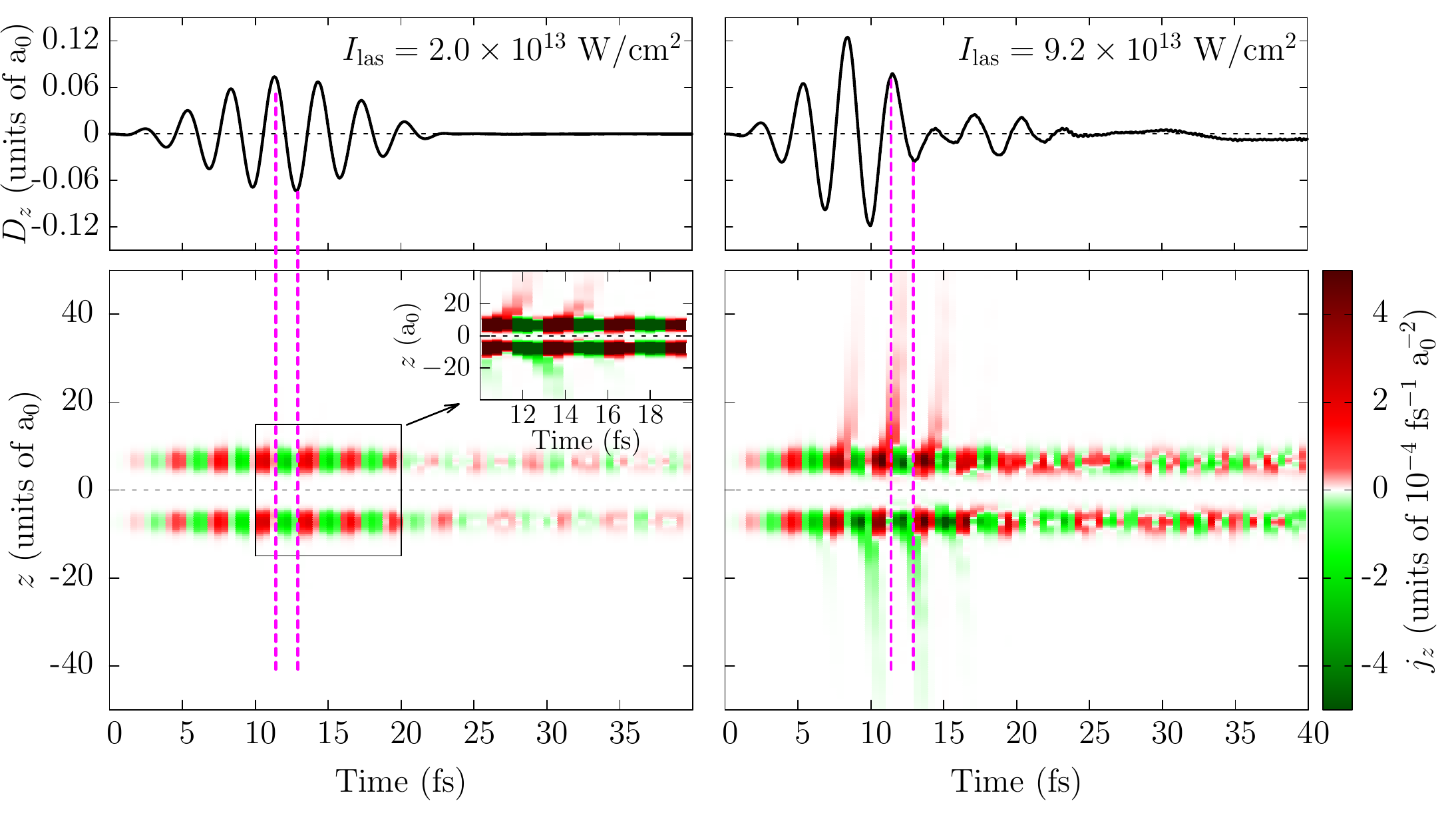}
 \caption{\label{fig:velo} Electronic dipole moment
   $D_z$ as a function of time (top row), and current density $j_z$ as a function of time and $z$ coordinate
 (bottom row, 2D density map) of C$_{60}$
   along laser polarization axis ($z$ direction), after irradiation by
   a laser pulse with $\omlas=1.36$~eV, $T_{\rm pulse}=24$~fs, and with two different intensities $\llas=2.0 \times
   10^{13}~\wcmq$ (left column) and $\llas=9.2 \times 10^{13}~\wcmq$
   (right column). Positive $j_z$ are depicted in red (or
   gray), and negative $j_z$ are in green (or light gray). For a given
   color (or type of gray), the brighter the shade, the smaller the
   absolute value of $j_z$.  The inset in the left bottom panel
   magnifies the map between 10 and 20 fs with a scale in current
   density divided by 100.}
\end{figure*}

To better visualize the ponderomotive motion, we show in the lower
panels of Fig.~\ref{fig:velo} the current density $j_z$ along the
  laser polarization axis $z$.  
Two intensities are considered, one
at the moderate side with $\llas=I_0=2 \times 10^{13}~\wcmq$ (left
column) and another one at higher intensity of $\llas=4.6\, I_0=9.2
\times 10^{13}~\wcmq$ (right column).  The current density is plotted
in a 2D density map as a function of time (horizontal axis) and space
coordinate $z$ (vertical axis), with positive $j_z$ in red (or gray)
and negative $j_z$ in green (or light gray).  It is also instructive
to compare these maps with the time evolution of the dipole moment defined in 
Eq.~(\ref{eq:dip}) which is plotted in the top panel of the figure. Note
that the dipole moment and the current density are related by the
continuity equation which reads~:
\begin{equation}
 \frac{\partial D_{z}}{\partial t}=\int_{z} \textrm dz~j_{z}
  \;.
\label{eq:curr}
\end{equation}
The sign of the time derivative of $D_z$ is thus equal to the one of
the dominant part of $j_z$. 

One observes successive fringes in $j_z$ connected to the
change of sign of the derivative of $D_z$ or, in other words, to the
oscillations of $D_z$ in time, as exemplified by vertical dashes in a
case when the sign is negative.  For the highest laser intensity
(right column of Fig.~\ref{fig:velo}), a sizable backflow of the
current density occurs during pulse duration, especially between 5 and
15 fs, where the field amplitude is maximal. For instance, in the time
interval indicated by the two vertical dashed lines, the majority of
electrons possess a negative $j_z$ (they are thus pulled away from the
$\csixty$), while a non-negligible amount exhibits a positive $j_z$,
which means that they are pulled back towards $\csixty$ and that
recollision is possible. At the lowest $I_\mathrm{las}$, this
backflow still exists but is much weaker, see inset in the bottom
left panel for which the current density scale has been divided by 100
to allow the visualization of this weak counterflow.  The
  amplitude of this quiver motion can be estimated from a purely
  classical model as $l_\mathrm{q}=\alpha E_0/\omega^{2}$ where
  $\alpha$ is the field enhancement factor (here about 1.05). This
  yields $l_\mathrm{q}\sim12~\bohr$ and $\sim25~\bohr$ at the low
  and the high $I_\mathrm{las}$ respectively. These classical
  $l_\mathrm{q}$ agree well with the amplitudes one can read
  off from the current density maps. The large ponderomotive oscillations terminate as soon as the external
field dies out.  The further evolution still shows a succession of
positive and negative fringes of $j_z$, but electrons in the $z>0$ and
the $z<0$ regions do not oscillate in phase anymore.  This is
particularly visible for the highest $I_\mathrm{las}$ above 20~fs.

\subsection{Time-resolved analysis}
\label{sec:timeanaly}
As mentioned above, 
in addition to the high-energy cut-off, we see modulations of the PES
within the broad plateau for the two cases with the higher intensities
in Fig.~\ref{fig:PES_Intens}. Similar structures 
have been experimentally observed in the photoemission spectra of rare gases (i.e., argon
atom~\cite{PauG01}) ionized by strong infrared laser
pulses. These structures have been interpreted as 
interference effects from several electron trajectories \CZG{generated either in the same optical cycle or in the subsequent optical cycle}, leading to the same final states ~\cite{Lew94}. 
In particular, in this work, two types of trajectories 
have been identified, labeled as ``short\rq{}\rq{} and ``long\rq{}\rq{}
trajectories, due to their different excursion times.
Within the three-step model, the electrons are 
born close to the field maxima, where the tunnel ionization probability is maximal. 

\CZG{Since we \MD{focus on} electron emission in the forward direction, 
we have chosen, for the time-resolved analysis, \MD{instants based on a classical picture of the
electron emission. More precisely, the ``birth'' times of the electrons, denoted by 
$t_{\rm b1/2/3}$,  are taken} at the three largest 
maxima of the electric force, \MD{that is at respectively}
2.75, 3.75, and 4.75~$t_\mathrm{oc}$, where $t_\mathrm{oc}=3$~fs is the single optical cycle. \MD{We have indicated these birth times} 
in Fig.~\ref{fig:e136TRPES}(a). \MD{We then correlate $t_{\rm b1/2/3}$ with the detection time $t_{\rm f1/2/3}$.} Since the (fastest) electrons generated at $t_{\rm b}$
need time to reach the boundary where they are detected, we take into account a time delay $\Delta t$ which consists of \MD{two terms. The first one is 
the ``return'' time for the electron to recollide with the target and is about $0.75t_\mathrm{oc}$. The second one is the time
for electrons emitted from the rescattering site to travel to the detection points $R_{b}=90~\mathrm{a}_{\rm 0}$ near the boundary of the simulation box.
This ``traveling'' time is $(R_{b}-R)$/$\sqrt{2E_\mathrm{kin}}$}
for electrons recorded with kinetic energy $E_\mathrm{kin}$. \MD{All in all, we have used~:}
\begin{subequations}
\label{eq:tfin}
\begin{eqnarray}
t_f&\approx& t_{b}+\Delta t,\\
\Delta t&=&0.75t_\mathrm{oc}+(R_{b}-R)/\sqrt{2E_\mathrm{kin}}.
\end{eqnarray}
\end{subequations}
\MD{For $t_{\rm b1}$, we used $E_\mathrm{kin}= 70$ eV leading to $\Delta t=1.04~t_\mathrm{oc}$, while for $t_{\rm b2}$ and $t_{\rm b3}$,
we took $E_\mathrm{kin}= 135$~eV and then obtained $\Delta t=0.96~t_\mathrm{oc}$. Plugging these numbers in Eqs.~(\ref{eq:tfin}), 
we got $t_{\rm f1/2/3}=$~3.79, 4.71, and 5.71~$t_\mathrm{oc}$ respectively, as symbolized in Fig.~\ref{fig:e136TRPES}(b).}

\MD{The bottom panel of Fig.~\ref{fig:e136TRPES} presents three different PES, note that each one has been analyzed from $t=0$ up to one of the
``final'' times introduced above. 
\begin{figure}[htbp]
 \centering
 \includegraphics[width=\linewidth]{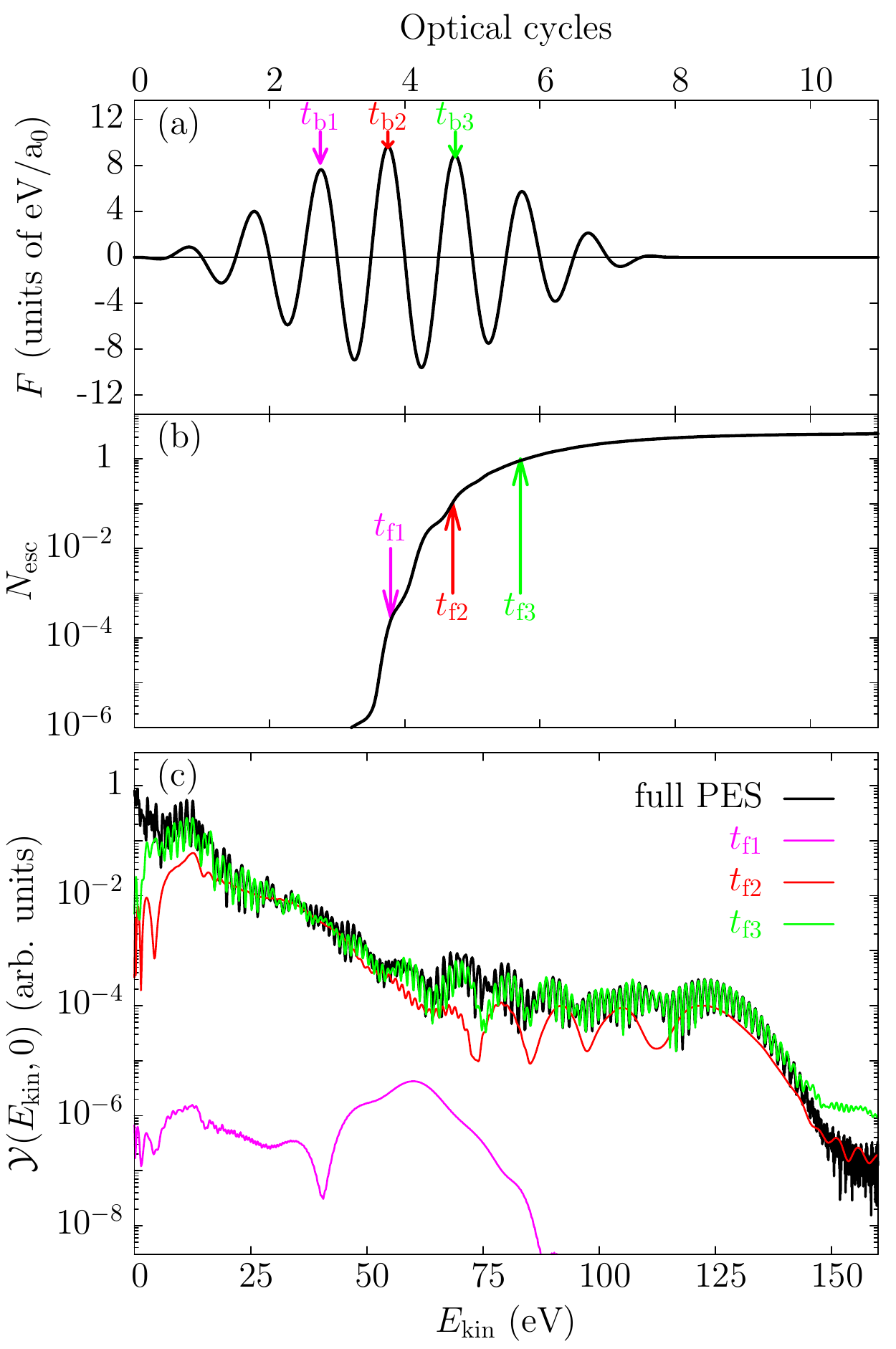}
 \caption{\label{fig:e136TRPES} Time-resolved analysis
   of photo-emission from $\csixty$ irradiated by a laser pulse
   polarized in the $z$ direction, with $\omlas=1.36$~eV, 
   $T_{\rm pulse}=24$~fs and 
   $I_\mathrm{las}=1.6\times{10}^{14}~\wcmq$.  (a) Time
   evolution of the force $F$ acting on electrons, with 3 ``birth'' times indicated by arrows (see text for details).
(b) Time evolution of the total ionization
     $\nesc$, with 3 ``final'' times indicated by arrows and defined in Eqs.~(\ref{eq:tfin}). 
 (c) Full PES
     (black) and PES from data accumulated up to the 3 different
     final times.}
\end{figure}
In addition, the full PES (black curve) obtained
at the end of the laser pulse is shown. We can therefore progressively observe how the full PES builds up in time.}
It should be noted that this analysis is strictly valid only for the high-energy electrons, since low-energy electrons arrive 
later, and are thus not accounted for at the corresponding 
final times $t_{\rm f1/2/3}$.}

As a first result, we see that the spectra at $t_{\rm f1}$ is 
about two to three orders of magnitude smaller than 
the spectra at later times (also see the corresponding ionization in Fig.\ref{fig:e136TRPES}(b)), showing that 
these contributions are negligible. Indeed,
at the field maximum \MD{indicated by $t_{\rm b1}$}, on the rising part of the pulse \MD{envelop}, the 
field strength is not yet sufficient for an efficient 
tunnel ionization. 
The major part of the spectrum, containing the main structures, 
is obtained for the spectrum at $t_{\rm f2}$, which corresponds \MD{to a birth time $t_{\rm b2}$ at}
the highest field value of the pulse. It shows a clear plateau, 
with distinct modulations. If one interprets these structures 
as interferences from two electron trajectories, 
it would correspond to a relative phase 
of $\delta S=E_\mathrm{kin} \delta t$~\cite{Mil06} with $\delta t$=0.41 fs, 
well below the optical period $t_\mathrm{oc}$= 3 fs.
This value of $\delta t$, about one seventh of $t_{\rm oc}$,
is in the same order as deduced by an extended schematic model for recent laser-nanotip interactions, estimated to about 
one sixth in this case~\cite{Kru12a}.
Within this interpretation 
of interference of different electron trajectories, 
the findings of the numerically calculated spectra
at $t_{\rm f2}$ would correspond to ``intracycle'' interferences, 
i.e., interference that stems from electron trajectories 
originating during the same optical period. 
When comparing to the spectrum at $t_{\rm f3}$, one sees that the 
additional changes with respect to that at $t_{\rm f2}$ 
are minor \MD{at the side of the extension of the plateau}, however, one striking difference 
can be observed: the appearance of high-frequency modulations 
in the spectra. These modulations correspond to peaks separated 
by the central frequency $\omega_\mathrm{las}$. 
From a temporal point of view, these structures are created by
interferences of electron trajectories that are born 
at times separated by the optical period, and can thus be 
identified as ``intercycle'' interferences. These are commonly observed 
for longer pulses, where several field peaks have comparable maxima, as it is in the present case.  

When compared to the full spectra, 
one sees slow convergence for the low-energy electrons. 
This can be understood by the fact that these electrons, 
due to their slower speed, only arrive at delays much larger 
than $\Delta t$. This systematic effect is clearly 
visible in all time-resolved spectra.

To summarize, we have analyzed the time evolution of the spectra
obtained by the fully numerical TDDFT calculations 
by choosing particular detection times to separate 
as fully as possible 
the contributions from the different field maxima.
While not claiming to have unambiguously identified the 
observed structures, we have shown that their 
characteristics, both in their time evolution as well as 
with respect to their
frequency fingerprints, are consistent with the 
picture of interfering intra- and intercycle trajectories. 

\section{Conclusions}
\label{sec:conclusion}

In this paper, we have analyzed the electron dynamics of laser-excited
$\csixty$ as a model case for the generation of
high-energy electrons from a carbon tip. We have explored the
response of the system to laser fields of various intensities for
  a frequency in the infrared domain which leads, at high intensities,
to significant ponderomotive effects. To analyze such a complex
dynamics, we go beyond single-active-electron approaches and use
time-dependent density-functional theory propagated in
real time and computed on a spatial grid. This approach allows, in
particular, a proper description of collective electronic motion as
well as a detailed analysis of photo-electron spectra (PES) and
  photo-angular distributions (PAD).  One of the main numerical challenges in the
present investigation was the large box size required to account for
the huge pathway of the ponderomotive motion of the electrons.
To make that feasible in a quantum mechanical
framework, we use a spherical jellium approximation for the ionic
background and handle the dynamics on a cylindrical grid.
 
As a major result, 
we have shown that the recollision regime can be reached for
strong, but realistic, laser intensities.  
We find the establishment
of a plateau stretching out to
very high kinetic energies, e.g. to about 125 eV for $I=1.6 \times10^{14}$ W/cm$^2$, which is interpreted
using the well-known recollision mechanism and 
illustrated by using a map of the time dependence of the
current distribution.  
The cut-off of the plateau is shown to follow the semiclassical 
model based on the three-step model, \MD{with a field which is enhanced by about $10 \%$ in our simulations.}
A detailed time-resolved analysis of the PES
demonstrates how the high-energy plateau is generated successively
during the laser pulse.
In particular, we have shown that the observed structures and their 
time evolution stemming from different peaks of the field are consistent 
with the picture of intra- and intercycle interferences. 
As far as the angular distribution is concerned, 
one of the most promising results of the presented study is
the strong focusing of the electrons in the forward direction, 
especially for the high energy electrons provided by the recollision process.   
This is an important feature in the context of using carbon nanotubes as future
sources of collimated electron beams for time-resolved diffraction
experiments. 
\newpage

\section*{Acknowledgments:}
C.-Z.G. thanks the financial support from China
Scholarship Council (CSC) (No.  [2013]3009). We thank Institut
Universitaire de France, European ITN network CORINF and French ANR
contract LASCAR (ANR-13-BS04-0007) for support during the realization
of this work. It was also granted access to the HPC resources of
CalMiP (Calcul en Midi-Pyr\'en\'ees) under the allocation P1238, and
of RRZE (Regionales Rechenzentrum Erlangen).\\

\bigskip
\bibliographystyle{apsrev}
\bibliography{lib_sfe.bib}

\end{document}